\documentclass[aps,prd,twocolumn,groupedaddress]{revtex4}
\usepackage{graphicx}
\usepackage{dcolumn}
\usepackage{bm}
\begin{document}

\title{First order phase transition in the quark matter}

\author{Masaharu Iwasaki}
\email{miwasaki@cc.kochi-u.ac.jp}
\affiliation{Department of Physics, Kochi University, Kochi 780-8520, Japan}

\date{\today}

\begin{abstract}
We investigate chiral phase transition of the first order in the quark matter. Using the Nambu-Jona-Lasinio model, an equation of state of the quark matter which is similar to the van der Waals' one is obtained. Moreover the specific heat and the compressibility are calculated. It is shown that they are enhanced in the symmetry broken phase, in particular diverge near the tricritical point.
\end{abstract}


\maketitle

\section{Introduction}

Thermodynamics of the quark matter have been discussed recently by many authors \cite{RW01}-\cite{S04}. One of the reasons comes from the possibility of the quark-gluon plasma (QGP), which lies in the chirally symmetric phase. It is expected that the QGP may be realized in high energy heavy-ion collisions at CERN Large Hadron Collider (LHC) and at the BNL Relativistic Heavy Ion Collider (RHIC). On the other hand the usual hadronic matter is in the chiral symmetry broken phase. Therefore it is of fundamental interest in the study of the chiral phase transition in the quark matter.

The chiral dynamics of the quark matter has been studied mainly in the framework of the Nambu-Jona-Lasinio (NJL) model or QCD-like theory \cite{AY89}-\cite{KMT01}. According to them, the phase transition is second order in the three-flavor quark matter. On the other hand, it is second order at low density and first order at high density in the two-flavor case. The critical point between them is called a tricritical point \cite{SRS98}-\cite{HJSSV98}. This behavior of the phase diagram was first shown by the NJL model \cite{AY89} and then confirmed by the QCD-like model \cite{BCDGP90}-\cite{KMT01}. It is very interesting for us to have the first order transition. The present author has calculated the latent heat appeared in the transition and pointed out that it may be used as a signal of the chiral transition \cite{IK03}.

The classical liquid-gas phase transition is the typical first order one. There exists large fluctuation near the tricritical point. This phenomenon is known as ``opalescence" first observed by T.Andrews in the nineteenth century \cite{A69}. It is generally expected that there is also large fluctuation in the quark matter and some physical quantities are enhanced near the point. These may be other signals of QGP. It is the purpose of this paper to calculate such quantities, the specific heat and the compressibility and investigate their behaviors near the critical temperature, in particular the tricritical point. In the course of this work we will obtain an equation of state of the quark matter, which is similar to the van der Waals' one.

The outline of the present paper is as follows. In the next section, we introduce the NJL model and calculate thermodynamic potential with the use of the mean field approximation. In Sec.{\rm III}, the specific heat and the compressibility are derived from the potential. The numerical calculation of the thermodynamic quantities is carried out in Sec.{\rm IV}. Section {\rm V} is devoted to discussions and summary with the help of the Landau theory of the phase transition.

\section{Thermodynamic potential}

In this section we will derive the thermodynamic potential of the quark matter in the frame of the NJL model \cite{HK94}. Our quark matter is supposed that the number of flavors is $N_f=2$ and that of colors is $N_c=3$. It is known that there is the chiral transition of the first order and the tricritical point exists in this case \cite{AY89}. The Lagrangian density of the NJL model is defined by
\begin{equation}
{\cal L}=\bar{\psi}(i\partial -m)\psi+g[(\bar{\psi}\psi)^2+(\bar{\psi}i\gamma_{5}{\bf \tau}\psi)^2 ],
\end{equation}
where $m$ is a current $u$- and $d$-quark mass and neglected in this paper for the convenience (chiral limit). The $g$ in this equation denotes a coupling constant of the quark-quark interaction and ${\bf \tau}$ is the Pauli matrix for flavors. This Lagrangian leads to the Hamiltonian density,
\begin{equation}
{\cal H}=\bar{\psi}(-i{\bf \gamma}\cdot \nabla -m)\psi-g[(\bar{\psi}\psi)^2+(\bar{\psi}i\gamma_{5}{\bf \tau}\psi)^2 ].
\end{equation}

Here we introduce two mean fields: $\sigma_1\equiv \langle\bar{\psi}\psi\rangle$ and $\sigma_2\equiv \langle\bar{\psi}\gamma^{0}\psi\rangle$. The expectation values represent thermodynamic averages and determined from the variational principle later. Then we obtain the linearized Hamiltonian density,
\begin{equation}
{\cal H}_{mf}=\bar{\psi}(-i{\bf \gamma}\cdot\nabla+M+\frac{g}{N_c}\sigma_{2}\gamma^{0})\psi + G\sigma_{1}^2 -\frac{g}{2N_c}\sigma_{2}^2,
\end{equation}
where $G\equiv (4N_{c}+1)g/4N_{c}$ is the renormalized coupling constant and $M$ is the effective quark mass defined by
\begin{equation}
M=m-2G\langle \bar{\psi}\psi\rangle.
\end{equation}

Now let us introduce the thermodynamic potential $\Omega(V,T,\mu)$ of our system by ($\beta\equiv 1/kT$)
\begin{equation}
\Omega=-kT\log {\rm Tr}\exp{[-\beta (H_{mf}-\mu N)]}\equiv V\omega(T,\mu),
\end{equation}
where $V$ and $T$ represent the volume and the temperature of our system respectively and $\mu$ is the chemical potential. The mean field Hamiltonian describes an assembly of free quasi-particles so that we can easily calculate the thermodynamic potential. Substituting Eq.(3) into this equation, we get
\begin{eqnarray}
\omega&=&E_{0}-\frac{N_{c}N_{f}}{\pi^2}T\int_{0}^{\Lambda}p^{2}\left[\log(1+\exp{-\beta(E-\mu_r)})\right. \nonumber\\
&&+\left.\log(1+\exp{-\beta(E+\mu_r)})\right]dp,
\end{eqnarray}
where a cut-off constant $\Lambda$ is introduced as usual and the first term $E_{0}$ is given by 
\begin{equation}
E_{0}\equiv G\sigma_{1}^2 -\frac{g}{2N_c}\sigma_{2}^2-\frac{N_{c}N_{f}}{\pi^2}\int_{0}^{\Lambda}p^{2}Edp.
\end{equation}
The quasi-particle energy and the renormalized chemical potential are defined by $E=\sqrt{{\bf p}^{2}+M^{2}}$ and $\mu_{r}=\mu-(g/N_{c})\sigma_{2}$ respectively. This potential $\omega$ plays a central role in the following discussion.

Lastly we must determine the mean fields $\sigma_1$ and $\sigma_2$. We take the variational principle of the thermodynamic potential: $\partial\omega/\partial\sigma_{1}=\partial\omega/\partial\sigma_{2}=0$. These equations lead to selfconsistency ones,
\begin{eqnarray}
\left\{
\begin{array}{l}
\sigma_{1}=-\frac{MN_{c}N_{f}}{\pi^2}\int_{0}^{\Lambda}\frac{p^2}{E}\left[ 1-n({\bf p},\mu_{r})-m({\bf p},\mu_{r})\right] dp ,\\
\sigma_{2}= \frac{N_{c}N_{f}}{\pi^2}\int_{0}^{\Lambda}\frac{p^2}{E}\left[ n({\bf p},\mu_{r})-m({\bf p},\mu_{r})\right] dp.
\end{array} \right.
\end{eqnarray}
The first equation is nothing but the gap equation in the BCS theory and the second determines the renormalized chemical potential.

\section{Specific heat and compressibility}

Now let us calculate thermodynamic quantities of the quark matter in order to study the critical behavior near the chiral phase transition. The specific heat and the compressibility reflect the large fluctuation, because they are proportional to the fluctuations of the entropy and the density respectively. The thermodynamic quantities are classified into two groups: extensive state quantities and intensive ones. As for the extensive state quantities, we use those per unit volume in this paper. For example, we consider rather the number density than the volume itself and discuss the thermodynamic potential per unit volume and so on. 

The differential of the thermodynamic potential reads to
\begin{equation}
d\omega=-SdT-\rho d\mu,
\end{equation}
where $S$ and $\rho$ denote the entropy density and the number density respectively. These quantities have discontinuities at the critical temperature of the transition \cite{IK03}. From Eq.(9), they are represented by
\begin{eqnarray}
S&=&-\frac{N_{c}N_{f}}{\pi^2}\int_{0}^{\Lambda} p^{2}\left[ n\log(n)+(1-n)\log(1-n)\right. \nonumber\\
&&\left.+m\log(m)+(1-m)\log(1-m) \right]dp, \\
\rho&=&-\frac{N_{c}N_{f}}{\pi^2}\int_{0}^{\Lambda} p^{2}\left[ n-m \right]dp.
\end{eqnarray}
This expression for the density is nothing but the mean field $\sigma_{2}$, which is a natural result from the definition. At this stage it should be noticed that we can obtain an equation of state for the quark matter. The pressure of the quark matter is given by $p=-\omega(T,\mu)$. On the other hand Eq.(11) shows that $\rho$ is a function of $T$ and $\mu$. From these equations we get the equation of state: $p=F(T,\rho)$.

Next let us consider the second derivatives of the thermodynamic potential, the specific heat and the compressibility. These quantities reflecting the fluctuation are supposed to be sensitive to the phase transition. They are defined by
\begin{eqnarray}
C&=&T\left( \frac{\partial S}{\partial T}\right)_{\mu}, \\
\kappa&=&\frac{1}{\rho}\left( \frac{\partial \rho}{\partial p}\right)_{T}.
\end{eqnarray}
However this expression for $\kappa$ is not useful for our calculation. It should be transformed into more tractable form. Using the formula of the Jacobian, it is rewritten to
\[
\left( \frac{\partial \rho}{\partial p}\right)_{T,\mu}=\frac{\partial(\rho,T)}{\partial(p,T)}=\frac{\frac{\partial(\rho,T)}{\partial(\mu,T)}}{\frac{\partial(p,T)}{\partial(\mu,T)}}
=\frac{\left(\frac{\partial \rho}{\partial\mu}\right)_T}{\left(\frac{\partial p}{\partial\mu}\right)_T}=\frac{1}{\rho}\left(\frac{\partial \rho}{\partial\mu}\right)_T,\nonumber
\]
with the use of $\rho=-\partial \omega/\partial \mu=\partial p/\partial \mu$. From this transformation, the compressibility turns out
\begin{equation}
\kappa=-\left( \frac{\partial \rho^{-1}}{\partial \mu} \right)_{T}.
\end{equation}
We will calculate these equations numerically and discuss the chiral phase transition in the next section.

\section{Numerical results}

In order to calculate the thermodynamic quantities considered in the previous section, the first task to do is to solve the self-consistency equations (8) with given values $\mu$ and $T$. The parameters of the NJL model are the same as those used in Ref.\cite{AY89}: $g=5.074\times 10^{-6}{\rm MeV^{-2}}, \Lambda=631{\rm MeV}, m=0$. As shown in the paper, the qualitative features of the phase transition do not depend on them. Then we have two solutions: $M=0$ and $M=M_{1}>0$. The former corresponds to the chiral symmetric phase and the latter to the broken phase. An example with $\mu=0.31{\rm GeV}$ (first order transition) is illustrated as a function of $T$ in Fig.1.
 \begin{figure}
 \includegraphics[width=\linewidth]{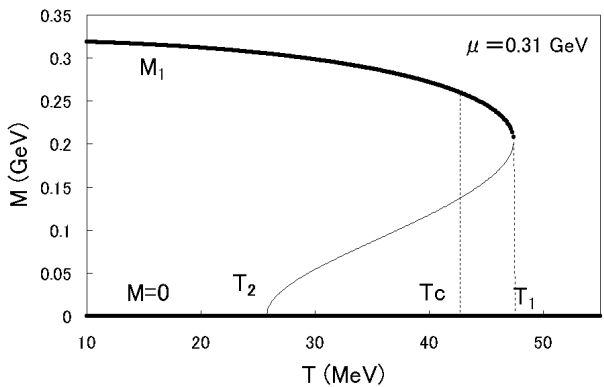}
 \caption{\label{fig:epsart}The effective mass $M$ as a function of the temperature at $\mu=0.31{\rm GeV}$. The critical temperatures $T_{c}$, $T_1$ and $T_2$ are defined in the text.}
 \end{figure}
The third solution appears in the range $T_{2}<T<T_{1}$. It is, however, always unstable so that we do not consider it. The stable state is realized by the solution with lower thermodynamic potential: The other solution corresponds to a metastable state. When the two solutions give the same value for the thermodynamic potential (degenerate), the phase transition occurs at the critical temperature $T=T_{c}$. Thus we have a phase diagram shown in Fig.2 where the critical temperature $T_{c}$ is drawn as a function of the chemical potential \cite{FK03}-\cite{K04}. 
 \begin{figure}
 \includegraphics[width=\linewidth]{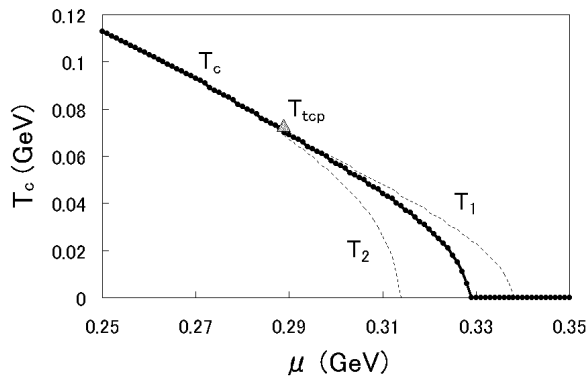}
 \caption{\label{fig:epsart}The critical temperature $T_c$ as a function of the chemical potential $\mu$. The tricritical temperature is denoted by $T_{tcp}$.}
 \end{figure}
The broken phase (hadron) is realized in the lower temperature ($T<T_{c}$) and the symmetric phase (QGP) in the higher one ($T>T_{c}$). The tricritical point is denoted by ${\rm T_{tcp}}$. It should be noted that the solution $M_{1}$ with the symmetry breaking is metastable even in the symmetric phase above $T_{c}$. This situation is known as supercooling. This metastable state exists in the range $T_{c}<T<T_{1}$ as shown by the broken line in Fig.1. On the other hand, the solution $M=0$ is also metastable in the broken phase ($T_{2}<T<T_{c}$). Hence the domain enclosed by the two dotted lines ($T_{2}<T<T_{1}$) has metastable states so that there would be large fluctuations in this domain. We call this domain transition region hereafter.

This circumstance is understood more clearly by studying the equation of state; the pressure is illustrated as a function of the inverse density as shown in Fig.3 with $T=30{\rm MeV}$.
 \begin{figure}
 \includegraphics[width=\linewidth]{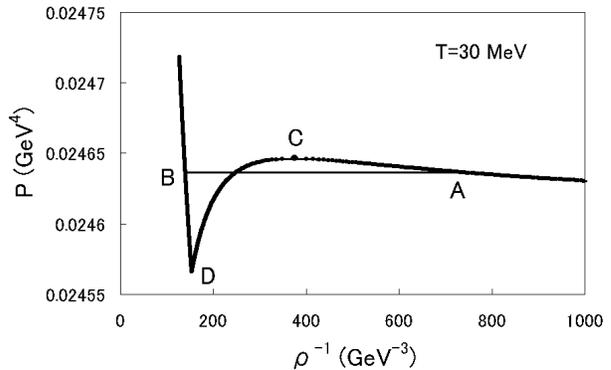}
 \caption{\label{fig:epsart}The equation of state of the quark matter at $T=30{\rm MeV}$. }
 \end{figure}
At once it is noticed that this equation of state is similar to the van der Waals' one. This behavior is based on the fact that an interaction between two particles in many-particle systems is attractive at large distance and repulsive at short distance. In the present case, our system has both the attractive force (quark-quark interaction) and the repulsive one due to the Pauli principle. Hence our equation of state in the Fig.3 is natural. The straight line AB where the chemical potential is constant is the Maxwell constraction. If the hadron phase is regarded as ``liquid" and the quark gas phase as ``gas", the present phase transition is opposite to that in the usual liquid-gas transition. The state ``A" (and outside) corresponds to the hadron (liquid) phase and the state ``B" (and inside) to the quark gas phase. The segment AC (BD) of the curve represents metastable state, which is called superheated liquid (supercooled gas). The remaining segment CD denotes unstable state. Therefore the quark matter jumps from A to B across the critical temperature $T_{c}$. This behavior is also understood by the solutions of the gap equation (8) as illustrated in Fig.4.
 \begin{figure}
 \includegraphics[width=\linewidth]{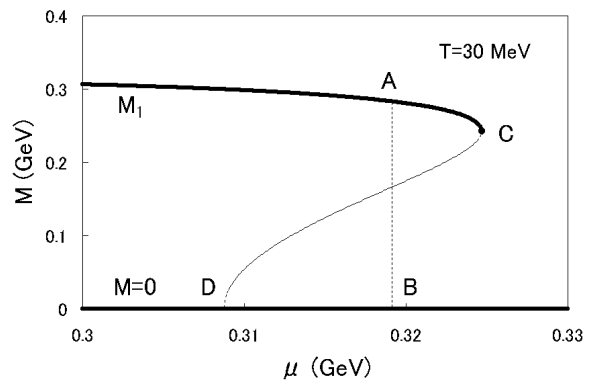}
 \caption{\label{fig:epsart}The effective mass $M$ as a function of the chemical potential at $T=30{\rm MeV}$. }
 \end{figure}
Here three solutions are drawn as a function of $\mu$ where four points (A, B, C and D) correspond to those in Fig.3.

Moreover we show the equations of state at three typical temperatures in Fig.5.  \begin{figure}
 \includegraphics[width=\linewidth]{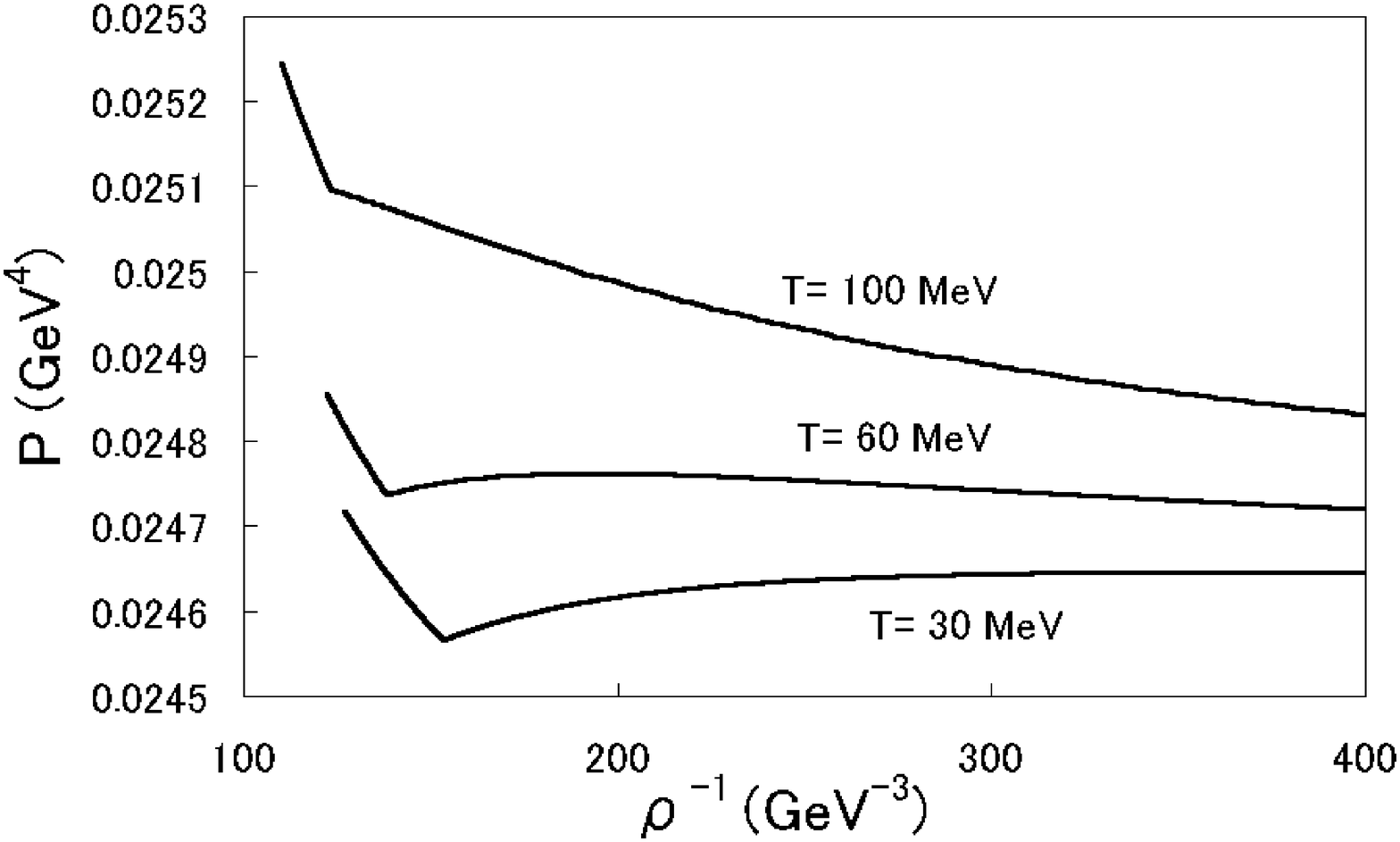}
 \caption{\label{fig:epsart}The equation of states at $T=100,~60,~30{\rm MeV}$.} \end{figure}
The equation of state at $T=100{\rm MeV}$ is monotonic since the phase transition is second order and the others are first order. The equation of state at $T=0$ has been already calculated in Ref.\cite{ARW98}.

Next let us calculate the specific heat and compressibility by difference approximation of Eqs.(12) and (14) respectively. They are shown in Figs.6 and 7 as a function of $T$ at three chemical potentials: The case of $\mu=0.25$ corresponds to the phase transition of the second order and the others to that of the first order. 
 \begin{figure}
 \includegraphics[width=\linewidth]{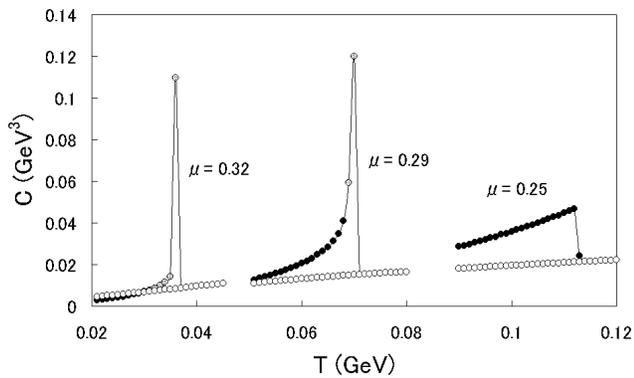}
 \caption{\label{fig:epsart}The specific heat of the quark matter as a function of the temperature at $\mu=0.25,~0.29,~0.31{\rm GeV}$.}
 \end{figure}
 
 \begin{figure}
 \includegraphics[width=\linewidth]{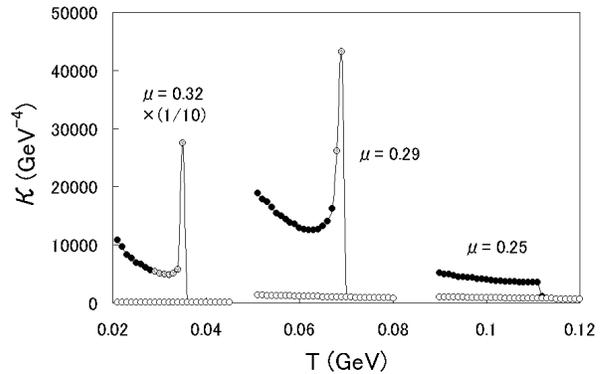}
 \caption{\label{fig:epsart}The compressibility of the quark matter as a function of the temperature at $\mu=0.25,~0.29,~0.31{\rm GeV}$. The values at $\mu=0.31{\rm GeV}$ are divided by $10$.}
 \end{figure}
The black circles in these figures represent the calculated values in the broken phase and the white ones in the symmetric one. Generally the specific heat in the broken phase is larger than that in the symmetric phase as well as the compressibility. The gray circles denote those in the metastable states. Although these are not realized within the mean field approximation, they may play an important role in the actual process. In fact soft modes will grow up in this transition region \cite{HK94}-\cite{HK85}. Namely the thermodynamic quantities of second derivatives are divergent near the temperature $T_{1}$, in particular the tricritical point. These properties may be used as a signal of the existence of the phase transition of the quark matter \cite{FO04}.

\section{Discussions and summary}

It has been shown that there appears the large enhancement of the specific heat and the compressibility near the tricritical point. These behaviors are due to the large fluctuation of the order parameter. Here we will clarify them in the context of the Landau mean field theory of phase transitions.

To this end we take $\sigma_{1}$ as an order parameter of the chiral phase transition and investigate the thermodynamic potential (6) as a function of $\sigma_{1}$. The other mean field $\sigma_{2}$ is regarded as a function of $\sigma_{1}$ through Eq.(8). If $\sigma_{1}$ is denoted by $\eta$, the thermodynamic potential is expanded as a power series of $\eta$ as follows:
\begin{equation}
\omega=\omega_{0}+A\eta^{2}+B\eta^{4}+D\eta^{6}+\cdots,
\end{equation}
where A, B and D are functions of $T$ and $\mu$. Moreover the $D$ is assumed to be always positive because of the stability of our system.

The order parameter should be determined by the stationary condition: $(\partial \omega/\partial \eta)=2\eta(A+2B\eta^{2}+3D\eta^{4})=0$. This equation has three solutions,
\begin{eqnarray}
\left\{
\begin{array}{l}
\eta_{0}\equiv 0, \\
\eta_{\pm}^2\equiv \frac{-B\pm \sqrt{B^{2}-3AD}}{3D},
\end{array}
\right.
\end{eqnarray}
where $\eta_{\pm}>0$ is assumed. Here our discussion is separated into two cases according to the sign of $B$: (a) $B>0$ and (b) $B<0$. In the case (a), the transition is second order. The stationary point with the minimum energy becomes
$\eta=\eta_{0}$ (for $A>0$) or $\eta=\eta_{+}$ (for $A<0$). Since the critical phenomena of the second order transition is well known, we will not consider this case hereafter. In later discussion we will restrict ourselves to the other case (b) that is the first order transition.

Then we have the following solutions depending on $A$, $B$ and $D$ or ($T$, $\mu$):
\begin{eqnarray}
\left\{
\begin{array}{ll}
(1)~~ (\eta_{0}),~ \eta_{+} &  ({\rm for}~ A<0) \\
(2)~~ \eta_{0},~ \eta_{+},~ (\eta_{-}) & ({\rm for}~ 0<A<B^{2}/3D) \\
(3)~~ \eta_{0}  &         ({\rm for}~ A>B^{2}/3D) 
\end{array}
\right.
\end{eqnarray}
The solution in the parentheses means an unstable solution. In the case (2) which is called the transition region, one of two extrema gives a stable state and the other a metastable state. The condition of the degeneracy ($\omega(\eta_{0})=\omega(\eta_{+})$ or $A=B^{2}/4D$) determines the critical temperature $T_{c}$. The $T_{1}$ ($T_{2}$) is determined by $A=B^{2}/3D$ ($A=0$). These circumstances are confirmed in the previous section by calculating our selfconsistency equations (8) directly. An example is shown in Fig.1 ($\mu=0.31{\rm GeV}$). It is seen that the above case (2) is realized in $26<T<47{\rm MeV}$.

Next we discuss the critical behavior of the specific heat near the temperature $T_{1}$. Since the condition of $T=T_{1}$ is given by $A=B^2/3D$, we can expand as $A-(B^{2}/3D)=a(T-T_{1})$ where $a$ is a positive function of $\mu$. Then the thermodynamic potential becomes
\begin{equation}
\omega=\omega_{0}-\frac{1}{27D^2}(X+B)(X-B)^2,
\end{equation}
where $X\equiv \sqrt{B^2-3AD}$. Differentiating with respect to $T$, we get
\begin{eqnarray}
S&=&S_{0}+\frac{a}{3D}(B-\sqrt{3aD(T_{1}-T)}), \nonumber\\
C&=&C_{0}+\frac{a\sqrt{3aD}}{6D}T(T_{1}-T)^{-1/2}.
\end{eqnarray}
It is seen that the entropy has finite gap across the temperature $T_{1}$. This generates the latent heat as discussed in Ref.\cite{IK03}. On the other hand, the specific heat diverges at $T_{1}$, which is consistent with the numerical result (Fig.6). As for the compressibility, we pay attention to the equation of state drawn in Fig.3. Noting that the compressibility can be rewritten as $\kappa=-\rho(\partial \rho^{-1}/\partial P)$, it is related to the derivative of the equation of state. From the curve we see the compressibility is enhanced in the metastable phase, in particular divergent at $T=T_{1}$, which is also seen in Fig.7.

The divergence of the specific heat and the compressibility can be also explained by using the thermodynamic potential. The temperature $T_{1}$ is defined by $B^2=3AD$, which gives
\begin{equation}
\omega=\omega_{0}+(A+B\eta_{+}^2)\eta_{+}^2+D\eta_{+}^6=\omega_{0}+D\eta_{+}^6.
\end{equation}
This cancellation of the second and fourth order terms leads to the large fluctuation of the order parameter.

In conclusion, we have developed the chiral phase transition of the first order in the framework of the NJL model. The quark matter discussed in this paper contains six kinds (two-flavor and three-color) of quark. Each of them has the same particle number, which means that our state is color singlet but not electric neutral. If this theory is applied to the bulk system, the charge neutrality condition would be important. Our calculation has been carried out using the mean field approximation, which does not take into account the pairing correlation, color superconductivity \cite{BL84}-\cite{ARW99}. This effect may be important for the high density and low temperature region and one of the future problems \cite{KKKN04}-\cite{T03}. We have obtained the equation of state of the quark matter, which is similar to the van der Waals' equation. Moreover the specific heat and the compressibility have been calculated. They are enhanced in the broken phase and are divergent in the metastable state, in particular at the tricritical point. It may be expected that this singular behavior appears in the expanding (cooling) process of the QGP generated in the high energy heavy-ion collisions. This expansion is like to the Big Bang and the inflationary expansion of the early universe \cite{KS81}-\cite{G81}. This singular behavior near the tricritical point may become a precursor of the chiral phase transition.

\begin{acknowledgments}
The author would like to thank Nuclear Theory Group at Kochi University for helpful discussions. He also thanks the Yukawa Institute for Theoretical Physics at Kyoto University. Discussions during the YITP workshop YITP-W-04-07 on ``Thermal Quantum Field Theories and Their Applications" were useful to complete this work.
\end{acknowledgments}

\end{document}